\documentclass[final,1p,times]{elsarticle}

\usepackage{graphicx}
\usepackage{amssymb}
\usepackage{float}
\usepackage{amsmath}

 \biboptions{sort&compress}

\makeatletter
\def\ps@pprintTitle{%
\let\@oddhead\@empty
\let\@evehead\@empty
\def\@oddfoot{}%
\let\@evenfoot\@oddfoot}
\begin{document}

\begin{frontmatter}


\title{Cavity Enhanced Interference of Orthogonal Modes in a Birefringent Medium}



\author{Kiran Kolluru$^a$ and  Subhasish Dutta Gupta$^{a,*}$}
\address[1]{School of Physics, University of Hyderabad, Hyderabad 500046, India\\  $^*$Corresponding author: sdghyderabad@gmail.com}

\begin{abstract}
Interference of orthogonal modes in a birefringent crystal is known to lead to interesting  physical effects (Solli et al., Phys. Rev. Lett. 91, 143906 (2003)). In this paper we show that the cavity with an intra-cavity rotator can enhance the mixing to the extent of normal mode splitting and avoided crossing depending on the orientation of the rotator with respect to the optic axis of the crystal. A high finesse cavity is shown to be capable of resolving small angles. The results are based on direct calculations of the cavity transmissions along with an analysis of its dispersion relation.
\end{abstract}


\end{frontmatter}


Interference of two or more resonances and their mixing has been one of the central themes of physics across it's various branches \cite{agarwal1984,sdggsa1985,raizen1989,spreeuw1990,gupta1995,armitage1998,reithmaier2004,rao2004,vahala2004,huang2009,gupta2015}. The participating modes can belong to different subsystems as in atom-field, exciton-photon and opto-mechanical systems etc. \cite{agarwal1984,raizen1989,armitage1998,reithmaier2004,rao2004,huang2009}. The modes can also be the independent modes of the same system. For example, they could be the clockwise and counter-clockwise modes of a fiber loop, spherical or disk resonators \cite{spreeuw1990,gupta1995,rao2005,rohde2007,chikkaraddy2013}. In metal/dielectric structures, these could be the surface plasmons at different interfaces, whispering gallery modes of metal-dielectric-metal micro and nano spheres \cite{gupta2009,peng2013}. One of the simple examples constitutes the mixing of orthogonal modes (ordinary and extraordinary waves) of a uniaxial birefringent crystal \cite{solli2003}. Obviously the cited examples constitute a very small subset of a very general and generic phenomenon of physics. Two different regimes of coupling, namely, weak and strong, have been identified and analyzed in detail \cite{wiersig2006}. The strong coupling regime is associated with an anticrossing or mode splitting of the resonant and near resonant modes \cite{wiersig2006,song2005,jansen2011}. The possibility  of increasing the lifetime of the interacting modes in a dissipative system has been highlighted \cite{song2005}. The mode splitting phenomenon has been observed in systems with large dissipation \cite{benz2013,lien2016}. There have been many other interesting applications of the normal mode splitting phenomena. These include fast and slow light \cite{solli2003,rao2004}, detection of  single-virus, single-molecules and nanoparticles \cite{vollmer2008,zhu2010}, and also optical sensing \cite{han2010}. Recently the splittings have been proposed as a tool for easier detection of the elusive Belinfante's transverse spin \cite{Belinfante1940,Berry2009} in a gap plasmon guide \cite{mukherjee2016}.
\par
In the context of the orthogonal modes in a birefringent crystal (BC), it was shown that a simple rotation of the detector horn can facilitate mixing of these modes \cite{solli2003}. The interference of these modes was shown to yield anomalous dispersion and superluminal light in an otherwise passive system. Note that usually anomalous dispersion is a characteristic of the absorption band of a dielectric. The effect reported by Solli et al. \cite{solli2003} can further be magnified in a cavity. In this letter we study a Fabry-Perot (FP) cavity containing a BC (with optic axis perpendicular to the cavity axis) and a polarization rotator (R). We choose the working point at a wavelength where both the ordinary and the extraordinary waves are resonant. We show that the combination of BC with R in the cavity can lead to a regime of normal mode splitting and anti-crossing for specific rotator orientations.
We present a vector formulation invoking the Jones matrices for the intra cavity elements to obtain the transmitted field and the dispersion relations. The numerical solutions of the transcendental dispersion equation are in full conformity with results for the intensity transmission. The fishnet in the $\theta-\lambda$ (rotator angle-wavelength) plane describing the peak transmitted intensities is explained by the closed-form solution of the dispersion equation. We also report an anti-crossing feature for specific points in  the ($\theta$-$\lambda$) plane. However, resolving the anti 
crossing features may be difficult because of the leakage-induced broadening of the split modes. Finally, we comment on the possible use of such a setup for measuring small angles, whereby the change in the rotator angles is mapped onto a well resolvable spectral separation in a high finesse cavity.
\par
\begin{figure}[h]
	{\includegraphics[width=\linewidth]{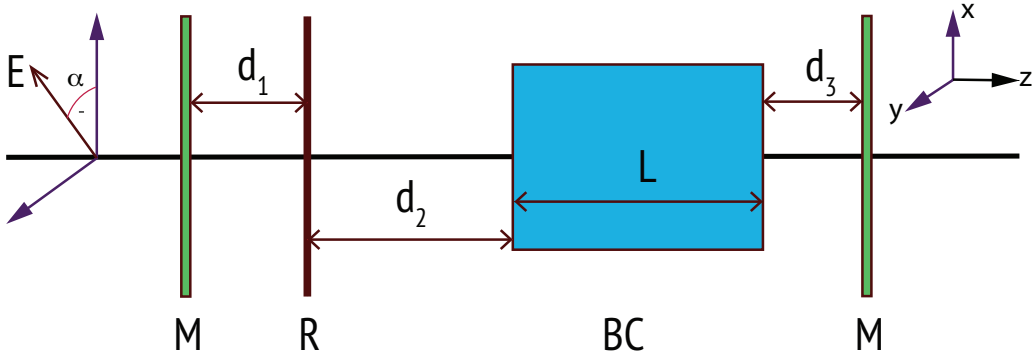}}
	\caption{(Color online) Schematics of the FP cavity with the mirrors M, the intra-cavity rotator R  and the birefringent crystal BC with optic axis along the $y$-axis.}
	\label{fig1}
\end{figure}

Consider the setup shown in Fig. \ref{fig1} comprising of a Fabry-Perot (FP) cavity with a intra-cavity rotator at $z=d_1$ and a birefringent crystal of length $L$. Let the cavity be illuminated by a plane polarized light propagating along the axis of the cavity ($z$-axis). The optic axis of the birefringent crystal is assumed to be oriented along $y$ direction. We assume the incident field to be polarized at an angle $\alpha$ with respect to the $x$-axis. For identical lossless mirrors (with ${\rho}^2+{\tau^2}=1$, where $\rho$ and $\tau$ are the  amplitude reflection and transmission coefficients of the mirrors, respectively) one can resort to the Jones matrix formalism to write the boundary conditions at $z=0$ and $z=d$ (where $d=d_{free}+L$ is the total length of the cavity with $d_{free}=d_1+d_2+d_3$ giving the free space propagation inside) as follows.
\begin{eqnarray}
\left(\begin{matrix}
E_{x}\\
E_{y}
\end{matrix}\right)_{z=0}&=& {\tau}{\left(\begin{matrix}
	E_{x}\\
	E_{y}
	\end{matrix}\right)_{in}}+{\rho}{\left(\begin{matrix}
	E'_{x}\\
	E'_{y}
	\end{matrix}\right)_{z=0}},\label{eq1}\\
\left(\begin{matrix}
E'_{x}\\
E'_{y}
\end{matrix}\right)_{z=d}&=&{\rho}{\left(\begin{matrix}
	E_{x}\\
	E_{y}
	\end{matrix}\right)_{z=d}},\label{eq2}
\end{eqnarray}
where  the subscript  `in' represents the input fields and the unprimed (primed) field amplitudes refer to the forward (backward) propagating waves, respectively. Using Eqs. (\ref{eq1}) and (\ref{eq2}) one can write down the transmitted amplitudes (denoted by subscript $tr$) as follows,
\begin{equation}
\left(\begin{matrix}
E_{x}\\
E_{y}
\end{matrix}\right)_{tr}={\tau}^2M_{f}(I-{\rho}^2{M_{total}})^{-1}\left(\begin{matrix}
E_{x}\\
E_{y}
\end{matrix}\right)_{in},\label{eq3}
\end{equation}
where $I$ is the $2\times2$ identity matrix and $M_{total}=M_b~M_f$ gives the round-trip Jones matrix for the cavity with $M_f$ ($M_b$) representing the Jones matrix for forward (backward) pass as follows \cite{hodgson2005,gupta2015,goldstein2016}
\begin{eqnarray}
M_b&=&M_{free}R(\theta)M_{BC},\label{eq4}\\
M_f&=&M_{BC}R(\theta)M_{free},\label{eq5}
\end{eqnarray}
with
\begin{equation}
\label{eq6}
M_{free}=e^{ik_0d_{free}}~I, ~~~~M_{BC}=Diag[e^{ik_0n_oL}, e^{ik_0n_eL}].
\end{equation}
In Eqs. (\ref{eq4})--(\ref{eq6}) $k_0$ is the free space propagation constant, $R(\theta)$ is the standard rotation matrix and $n_e$ ($n_o$) is the refractive index of the BC for the extraordinary (ordinary) wave. Introducing notations $\kappa$,~$\tilde{\delta}$ as
\begin{eqnarray}
\kappa&=&\dfrac{n_e+n_o}{n_e-n_o},\label{eq7}\\
\tilde{\delta}&=&k_0(n_e-n_o)L,\label{eq8}
\end{eqnarray}
the final form of $M_{total}$ can be written as 
\begin{equation}\label{eq9}
M_{total}=
e^{2ik_0d_{free}}e^{i\kappa\tilde{\delta}}\left(\begin{smallmatrix}
\cos^2{\theta} e^{-i\tilde{\delta}} -\sin^2\theta e^{i\tilde{\delta}}&-\sin2\theta \cos\tilde{\delta}\\
\sin{2\theta}\cos{\tilde{\delta}}& \cos^2\theta e^{i\tilde{\delta}}-\sin^2{\theta}e^{-i\tilde{\delta}}
\end{smallmatrix}
\right). 
\end{equation}          	          	                
It is clear that  without the rotator ($\theta=0^\circ$), the $x$ and $y$ components of the field experience phase shifts,
\begin{equation}
\label{eq10}
2k_0(d_{free}+n_{o}L)=2n\pi, ~~
2k_0(d_{free}+n_{e}L)=2m\pi, 
\end{equation}
\begin{figure}[h]
	{\includegraphics[width=\linewidth]{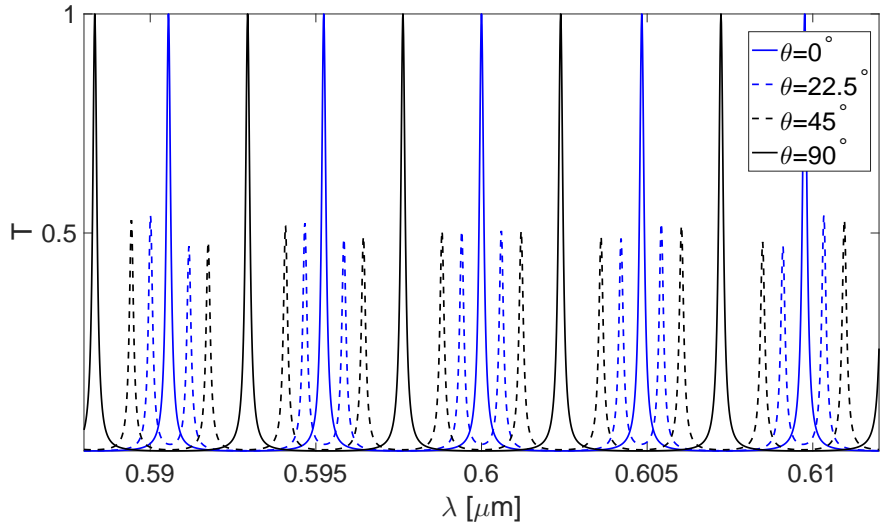}}
	\caption{(Color online) Intensity transmission $T$ as a function of $\lambda$ for varying rotator angles namely, $\theta=0^\circ$ (solid blue line), $22.5^\circ$ (dashed blue line), $45^\circ$ (dashed black line), $90^\circ$ (solid black line). Parameters are as follows, $\rho=0.95$, $L=10 ~\mu\rm m$, $d_{free}=15 ~\mu\rm m$, $n_e=2.28$, $n_o=2.25$.}
	\label{fig2}
\end{figure}
respectively, defining the ordinary and the extraordinary modes. In Eq. (\ref{eq10}) $m$ and $n$ are integers. We pick the cavity parameters such that both the ordinary and the extraordinary waves are resonant at the same wavelength, say at $\tilde{\lambda}_0$. To be specific, for a given length $L$ we choose $\tilde{\lambda}_0$ such that $\tilde{k}_0(n_e-n_o)L=(m-n)\pi=q\pi,~ \tilde{k}_0=2\pi/\tilde{\lambda}_0$. 
\par
\begin{figure}[t]
	\includegraphics[width=\linewidth]{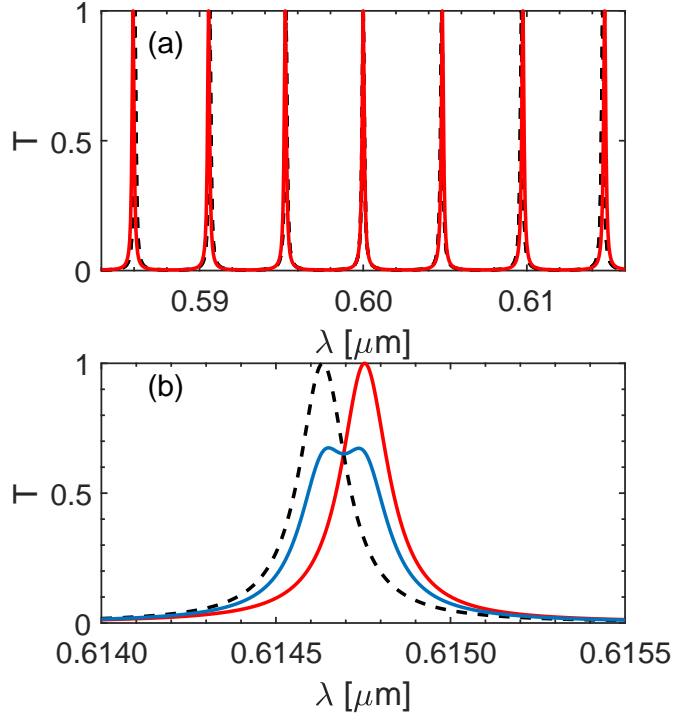}
	\caption{(Color online) (a) Intensity transmission $T$ as a function of $\lambda$ in the absence of rotator ($\theta=0^\circ$). The solid red line (dashed black line) shows the results for $x$- ($y$-) polarized incident field. (b) Intensity transmission $T$ as function of $\lambda$ for incident polarization angle $\alpha=45^\circ$ (blue line) in the absence of the intra-cavity rotator ($\theta=0^\circ$). For comparison we have shown the results for $x$-  and $y$- polarized incidence (Fig. \ref{fig3} (a)). The parameters are the same as in Fig. \ref{fig2}.}
	\label{fig3}
\end{figure}

In what follows we present the numerical results for the intensity transmission coefficient $T$ ($=T_x+T_y=|E_x|^2+|E_y|^2$ at the exit face) for an incident field with unit magnitude. We have chosen the following parameters for the numerical calculations: $\rho=0.95$, $L=10 \mu \rm{m}$, $d_{free}=15 \mu \rm{m}$, $n_o=2.25$ and $n_e=2.28$. For the chosen set, the two orthogonal modes are degenerate at $\tilde{\lambda}_0=0.6~ \mu \rm{m}$ (see Fig. \ref{fig3} (a) below). For an incident field polarized along the $x$-axis ($\alpha=0$), $T$ as a function of $\lambda$ for different orientations of the rotator is shown in Fig. \ref{fig2}. We have presented results for a range of wavelengths and for four different values of $\theta$, namely, $\theta= 0^{\circ}, 22.5^{\circ}$, $45^{\circ}$ and $90^{\circ}$. Note that the curves  for $\theta=90^{\circ}$ represent the orthogonal extraordinary modes. Further it can be seen that the mode splitting is asymmetric for modes that are not degenerate. A deeper insight can be obtained by analyzing the corresponding  dispersion relation which is presented below.
\par
In contrast, in absence of the rotator inside the cavity ($\theta=0^\circ$), even for tilted incident field (for finite $\alpha$) the splittings presented in Fig. \ref{fig2} are missing. For reference we have shown the ordinary (extraordinary) modes for $x$- ($y$-) polarized incident wave in absence of the rotator ($\theta=0^{\circ}$) in Fig. \ref{fig3} (a). The case for $\alpha=45^{\circ}$ is shown in Fig. \ref{fig3} (b), where the intensity transmission results from a superposition of the $x-$ and $y-$ components of the field. Clearly the peaks correspond to the resonances of the ordinary and extraordinary modes, which are not degenerate near $\lambda=0.6147~ \mu \rm{m}$. Thus the origin of splitting can be traced to the mode mixing by the rotator inside the cavity.
\par
We now show how a high finesse cavity can facilitate the measurement of small angles. To this end we have plotted the transmission spectra for a small rotator angle, say $\theta=0.5^{\circ}$ for three different values of the finesse. While, width of the split resonances mask the splittings for a low finesse cavity, for $\mathcal{F}=1000$ one can easily resolve the split modes. Thus an angular shift of $0.5^{\circ}$ can be translated to a spectral separation of $0.03$ n$\rm{m}$. Similar techniques have been adapted to count nanoparticles and to measure the presence of single virus \cite{vollmer2008,zhu2010}.
\begin{figure}[h]
	\centering
	{\includegraphics[width=\linewidth]{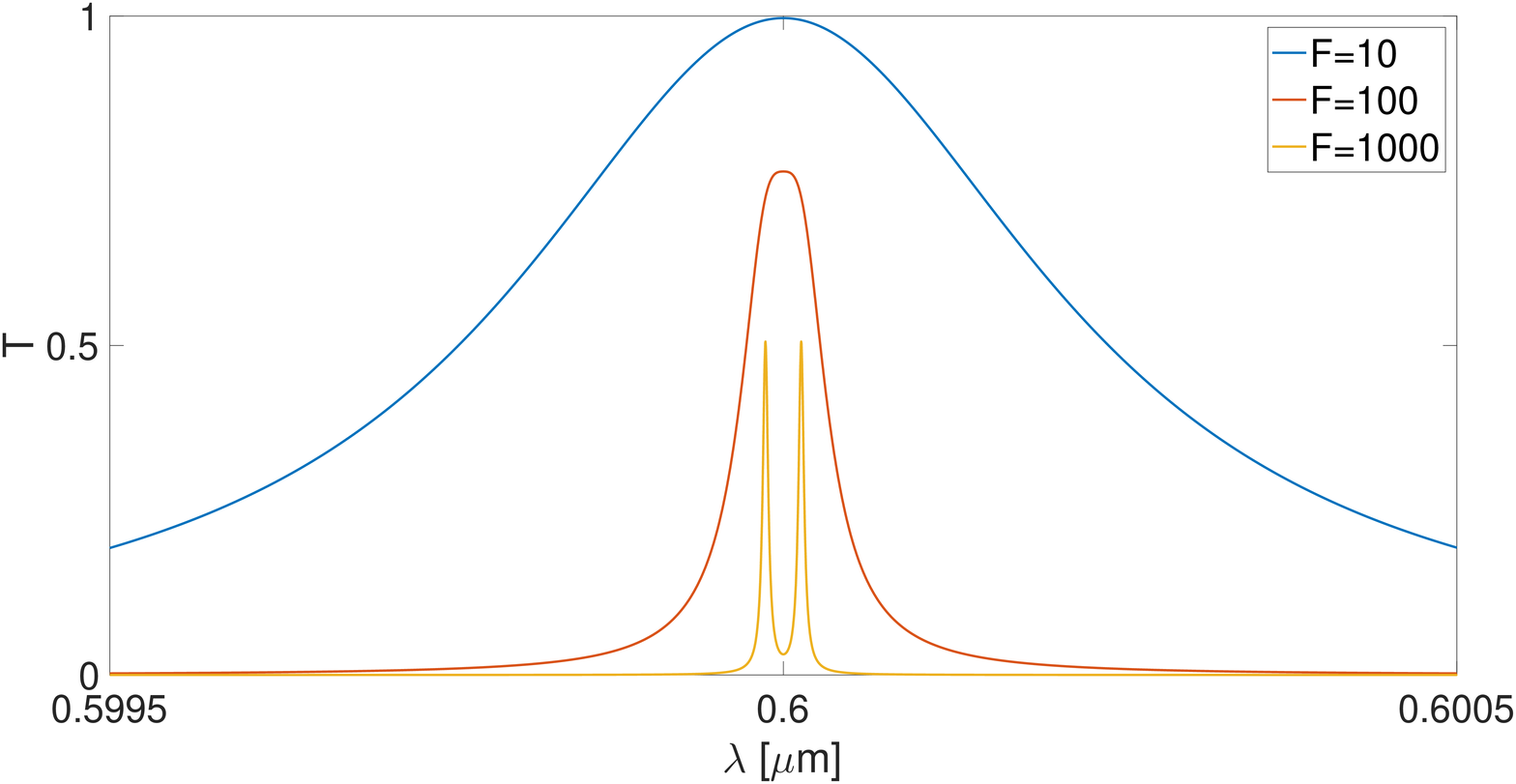}}
	\caption{(Color online) Intensity transmission $T$ as a function of $\lambda$ for three different values of finesse $\mathcal{F}= 10,~ 100~ \rm{and}~ 1000$ for $\theta=0.5^\circ$. Other parameters are the same as in Fig. \ref{fig2}.}
	\label{fig4}
\end{figure}
\par
\begin{figure}[t]
	\centering
	{\includegraphics[width=\linewidth]{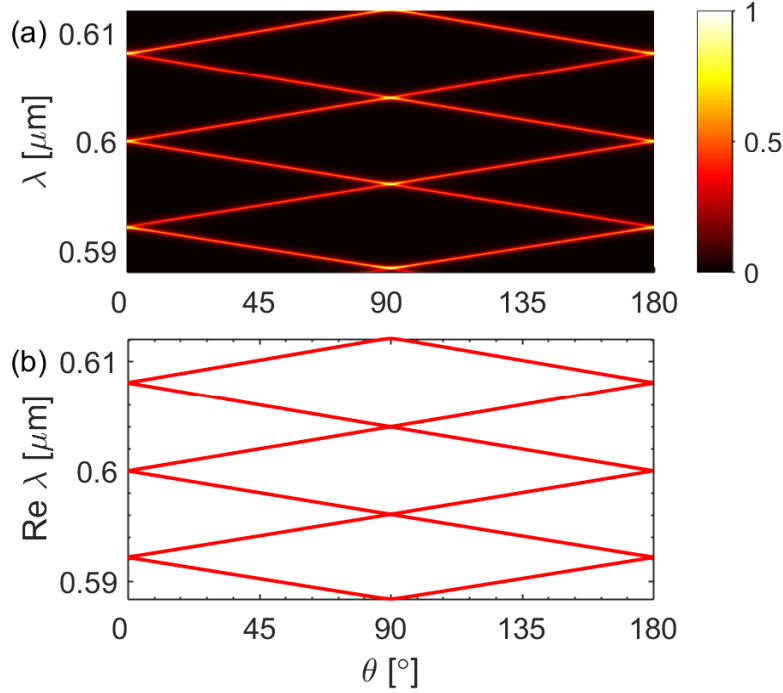}}
	\caption{(Color online) (a) Color plot of the intensity transmission $T$ as a function $\theta$ and $\lambda$. (b) Re($\lambda$) as a function of $\theta$ ( solution of the dispertion relation given by Eq. (\ref{eq12})). Except for $d_{free}=0$, remaining parameters are the same as in Fig. \ref{fig2}.} 
	\label{fig5}
\end{figure}

We next obtain and analyze the dispersion relation, explaining the mode splitting shown in Fig. \ref{fig2}. In order to make the analysis simpler (without any loss of generality or the essential physics), we set $d_{free}=0$. It may be noted that a finite $d_{free}$ introduces nominal changes in the dispersion diagram. So as to see the direct connection with the transmission results, the color plot of $T$ as functions of wavelength $\lambda$ and rotator angle $\theta$ is shown in Fig. \ref{fig5} (a). It is clear from Eq. (\ref{eq3}) that the poles of the transmission coefficient are given by the following equation
\begin{equation}
\label{eq11}
det(I-{\rho^2}{M_{total}})=0.
\end{equation}
The dispersion relation given by Eq. (\ref{eq11}) can be recast in the form
\begin{equation}\label{eq12}
\rho^{4}e^{2i\kappa\tilde{\delta}}-2\rho^{2} e^{i\kappa\tilde{\delta}}\cos2\theta\cos\tilde{\delta}+1=0 .
\end{equation}
Eq. (\ref{eq12}) does not allow any real solution for $\tilde{\delta}=k_0(n_e-n_o)L$ since $k_0$ (or the wavelength $\lambda$) is now complex. Note that the real part of $k_0$ gives the location of the resonances while the imaginary part gives the decay rates of the modes. With respect to our reference wavelength $\tilde{\lambda}_0$ (=0.6 $\mu\rm m$) we express $\tilde{\delta}$ as
\begin{equation}
\label{eq13}
\tilde{\delta}=\tilde{\delta} _0+\delta, ~~\tilde{\delta} _0=\tilde{k}_0(n_e-n_o)L,
\end{equation}
we look for complex solutions for the increment $\delta$. The results then can easily be mapped onto the real and imaginary parts of $\lambda$. Keeping in view the degenerate resonance at our reference wavelength, the tilde sign in Eq. (\ref{eq12}) can be dropped. Eq. (\ref{eq12}) represents a transcendental equation and was solved numerically using a root finding algorithm in MATLAB. The results for $Re(\lambda)$ is shown in Fig. \ref{fig5} (b). Note that while there is a splitting of the real part of $\lambda$, the imaginary part (giving the lifetimes) remains constant (given by $Im(\delta/(n_e-n_o)L)=0.0023~[\mu\rm m]^{-1}$). It is clear from a comparison of Fig. \ref{fig5} (a) and Fig. \ref{fig5} (b) that the dispersion analysis captures the splittings in the transmission data. In order to explain the seemingly linear behavior in Fig. \ref{fig5} (b), we approximate Eq. (\ref{eq12}) assuming small variation around our reference wavelength (i.e. $\cos \delta\sim1$). Then Eq. (\ref{eq12}) reads as
\begin{equation}\label{eq14}
\rho^{4}e^{2i\kappa \delta}-2\rho^{2} e^{i\kappa \delta} \cos 2\theta +1=0 .
\end{equation}
Eq. (\ref{eq14}) can be treated as a quadratic equation with respect to $\rho^2e^{i\kappa\delta}$ which can be solved to yield the root as follows
\begin{equation}\label{eq15}
\rho^2e^{i\kappa\delta}=\cos2\theta\pm i\sin 2\theta.
\end{equation}
The results given by Eq. (\ref{eq15}) can easily be translated to the wavelengths yielding the following
\begin{equation}\label{eq16}
\Delta \lambda=\pm\dfrac{\tilde{\lambda}^2_0(\theta\pm q\pi)}{\pi\kappa(n_e-n_o)L}+i \dfrac{\tilde{\lambda}^2_0\ln \rho}{\pi\kappa(n_e-n_o)L} .  
\end{equation}
From Eq. (\ref{eq16}) it is clear that while the real part is split in two distinct branches moving away from $\lambda= 0.6~\mu \rm m$, the lifetime is not split and is given by the same value of the imaginary part, which depends only on the system parameters. Integer $q=0,1,2\cdots$ yields the other branches. 
\begin{figure}[h]
	\centering
	{\includegraphics[width=\linewidth]{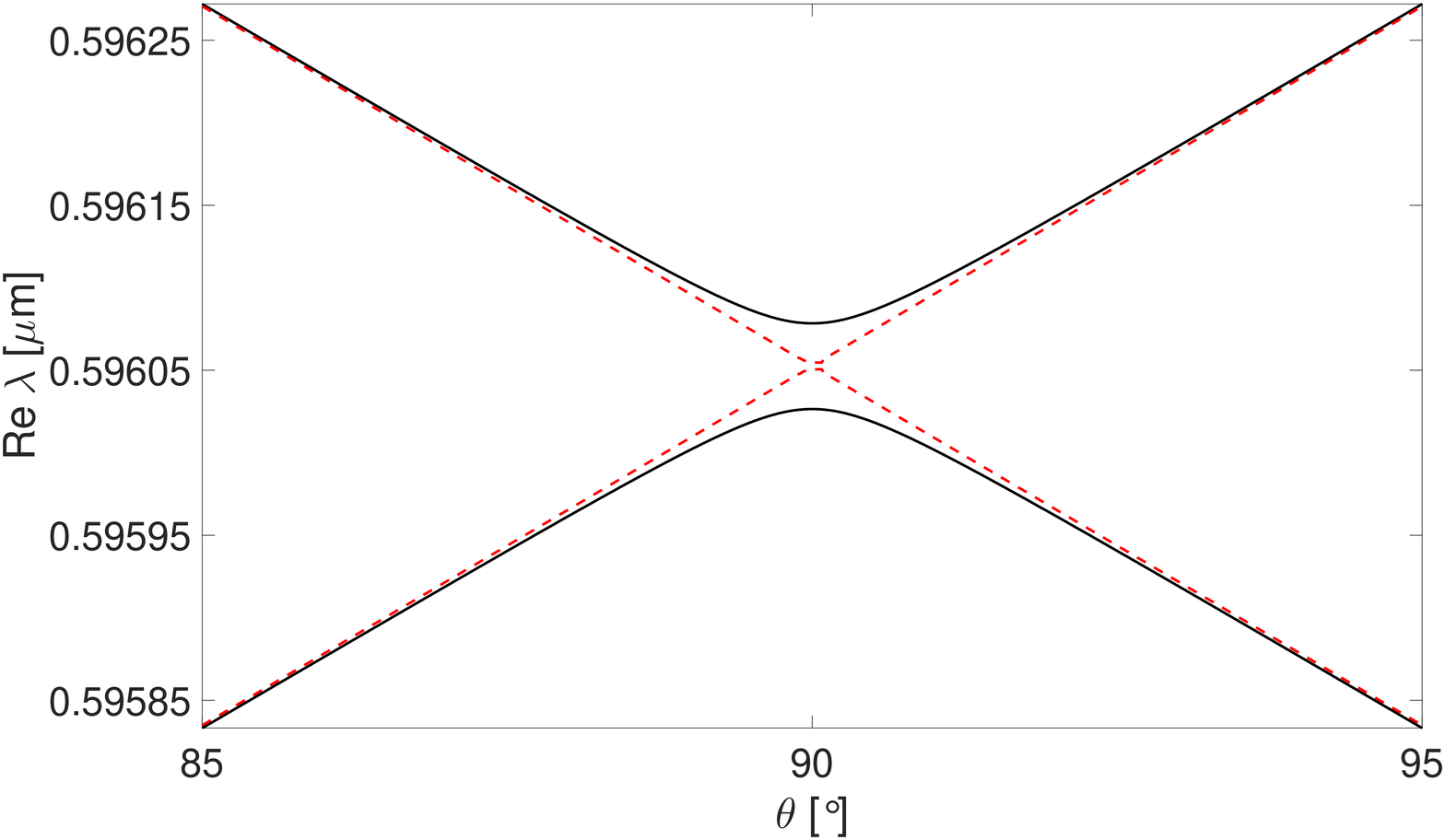}}
	\caption{ (Color online) Expanded portion of Fig. \ref{fig5} (b) giving the solution of the exact dispersion relation (\ref{eq12}) for Re$\lambda$ (solid black line). The dotted red lines gives the linear results of Eq. (\ref{eq16}). Parameters are as in Fig. \ref{fig5}.}
	\label{fig6}
\end{figure}
\par
Before we end this discussion, we show that there is the anti-crossing phenomenon associated with the crossing of the dispersion branches for $\theta=90^{\circ}$, say, near $\lambda=0.5961~\mu \rm m$. It is shown in Fig. \ref{fig6}, where we have plotted the solution for $Re(\lambda)$ of the dispersion Eq. (\ref{eq12}). The dashed lines in Fig. \ref{fig6} represent the crossing obtained from Eq. (\ref{eq14}) for reference. However, the tiny splitting at $\theta=90^{\circ}$ cannot be resolved because they are masked by the decay induced broadening of the split modes. Such anti-crossings are not there at the reference wavelength for $\theta=0^\circ$. In order to show this we assume $\delta$ to be purely imaginary ($\delta=i\delta^{\prime\prime}$), so that we have a solution for $Re(\delta)=0$ (i.e., $Re(\lambda)=0.6~\mu \rm m$). The existence of a real solution for $\delta^{\prime\prime}$ proves that there is no anti-crossing at $\lambda=0.6~ \mu \rm m$ with both the modes resonant and with no rotator in the cavity ($\theta=0^\circ$). With the aforesaid Eq. (\ref{eq12}) reads as follows
\begin{equation}\label{eq17}
\rho^{4}e^{-2\kappa \delta^{\prime\prime}}-2\rho^{2} e^{-\kappa \delta^{\prime\prime}} \cosh \delta^{\prime\prime} +1=0 .
\end{equation}
Eq. (\ref{eq17}) can be factorized to yield the two roots for $\delta^{\prime\prime}$ as follows
\begin{equation}\label{eq18}
\delta^{\prime\prime}=\dfrac{ 2\ln\rho}{\kappa \pm 1}.
\end{equation}
It is clear from Eq. (\ref{eq18}) that the lifetimes are split for the modes with the same wavelength $\tilde{\lambda}_0$. These correspond to the lifetimes of the uncoupled ordinary and extraordinary modes. In standard degenerate mode splitting phenomena the frequency split modes are characterized by the same lifetime, while lifetime splittings correspond to the modes with the same frequency \cite{gupta1995,lien2016}. Eqs. (\ref{eq16}) and (\ref{eq18}) reflect the same physics. Such solutions are not possible for the other intensity peaks shown in Fig. \ref{fig5}.
\par 
In conclusion, we have studied mode coupling between the ordinary and extraordinary waves in a birefringent FP cavity mediated by an intra-cavity rotator. The length of the cavity and the crystal are chosen so as to make both the ordinary and the extraordinary waves resonant at a given wavelength. The transmission spectra for an incident plane polarized light is shown to exhibit the normal mode splitting even for tiny angles $\theta$ of the rotator. We have presented a detailed analysis of the corresponding dispersion relation leading to solutions representing an intricate mesh in the $(\theta,\lambda)$ plane. Anti-crossing is shown for $\theta=90^{\circ}$ when a single pass through the cavity converts the nature of the participating modes (ordinary to extraordinary and vice versa). The results can find applications in precision measurement of small angles in a high finesse cavity, where angular variations are converted to spectral shifts.




\bibliographystyle{model1-num-names}
\bibliography{bibfile2.bib}







\end{document}